\documentclass[useAMS,referee, a4paper]{mn2e}  \usepackage{graphicx} 

 \title[CI  lines  as H$_2$ gas mass tracers]{The  CI lines  as  tracers  of
molecular gas, and their prospects at high redshifts}

\author[Papadopoulos, Thi, and Viti]{P. P. Papadopoulos,$ ^{1,2}$ W.-F. Thi,$^{3}$ and
S. Viti,$^{1,4}$\\
$^{1}$Department of Physics \& Astronomy, University College London,  
           London, WC1E 6BT, UK\\
$^{2}$ESA Astrophysics Division, Research and Scientific Support Department, ESTEC, 
              Postbus 299, 2200 AG Noordwijk, The Netherlands\\
$^{3}$Sterrenkundig Instituut ``Anton Pannekoek'', Kruislaan 403,
              1098 SJ Amsterdam, The Netherlands\\
$^{4}$CNR-Istituto di Fisica dello Spazio Interplanetario, Area di Ricerca di
Tor Vergata, via del fosso del Cavaliere 100, 00133, Roma, Italy }

\begin{document} 

\date{Received; Accepted}

\pagerange{\pageref{firstpage}--\pageref{lastpage}} \pubyear{2004}

\maketitle

\label{firstpage}

\begin{abstract}

We examine the fine structure lines $^{3}P_{1}\rightarrow $$^{3}P_{0}$
(492~GHz)  and $  ^{3}P_2\rightarrow $$^{3}P_1$  (809 GHz)  of neutral
atomic carbon  as bulk molecular gas  mass tracers and  find that they
can be good and on  many occasions better than $ ^{12}$CO transitions,
especially at  high redshifts. The notion  of CI emission  as an H$_2$
gas mass tracer challenges the long-held view of its distribution over
only a  relatively narrow  layer in the  CII/CI/CO transition  zone in
FUV-illuminated  molecular  clouds.   Past  observations  have  indeed
consistently pointed  towards a more  extended CI distribution  but it
was  only recently,  with the  advent of  large scale  imaging  of its
$^{3}P_{1}\rightarrow  $$^{3}P_{0}$  transition,  that its  surprising
ubiquity in molecular clouds has  been fully revealed.  In the present
work we show that under  {\it typical} ISM conditions such an ubiquity
is  inevitable  because  of  well known  dynamic  and  non-equilibrium
chemistry  processes   maintaining  a  significant   [C]/[$  ^{12}$CO]
abundance  throughout Giant  Molecular Clouds  during  their lifetime.
These processes are more  intense in star-forming environments where a
larger ambient  cosmic ray  flux will also  play an important  role in
boosting [C]/[$ ^{12}$CO].   The resulting CI lines can  be bright and
effective   H$_2$   mass  tracers   especially   for  diffuse   ($\sim
10^2-10^3\rm  \ cm^{-3}$)  gas while  in UV-intense  and/or metal-poor
environments  their  H$_2$-tracing  capability diminishes  because  of
large scale  CII production but nevertheless remains  superior to that
of $ ^{12}$CO.  The best place to take full advantage of CI's capacity
to trace H$_2$ is not in the low-$z$ Universe, where large atmospheric
absorption at  492 and 809~GHz precludes routine  observations, but at
high redshifts ($\rm z\ga 1$).

\end{abstract}


\section{Introduction}

The  use of  $  ^{12}$CO  and its  isotopologues  rotational lines  to
measure H$_2$  gas mass and physical  properties in galaxies  is now a
well-established  technique  (e.g.  Dickman  et  al.  1986;  Elmegreen
1989; Bryant  \& Scoville  1996) employed successfully  throughout the
local Universe  (z$\la 0.3$) (e.g.  Sanders, Scoville  \& Soifer 1991;
Young \& Scoville 1991; Solomon  et al.  1992; Downes \& Solomon 1998)
and at increasingly  high redshifts (e.g.  Brown \&  Vanden Bout 1991;
Downes et  al.  1992;  Omont et  al.  1996; Frayer  et al.   1998). In
high-z observations  usually only high-J CO  lines $\rm J+1\rightarrow
J$,  ($\rm J+1\geq  4$) are  accessible  for evaluating  the mass  and
physical conditions of H$_2$  gas, thus omitting cooler ($\rm T_{kin}<
30$  K)  and/or sub-thermally  excited  gas  ($\rm n<10^4\  cm^{-3}$).
Recently  observations   of  the   lowest  $  ^{12}$CO   J=1--0,  2--1
transitions in  several high redshift objects have  been made possible
with the Very Large Array  (e.g. Carilli, Menten, \& Yun 1999; Carilli
et al.  2002a, 2002b) and  in some cases thermalized transitions up to
fairly  high  rotational levels  (e.g.   J=5--4)  have been  observed.
However, differential  lensing can complicate  standard interpretation
of  line  ratios by  preferentially  amplifying  the  emission of  the
usually more spatially confined warm  and dense gas (e.g.  Blain 1999)
yielding  the  false impression  that  such  gas  properties (and  the
thermalization of high-J CO  transitions) characterize the bulk of the
molecular  gas in such  systems.  In  one case  of an  unlensed high-z
starburst large scale sub-thermal  excitation of $ ^{12}$CO J=5--4 has
been inferred (Papadopoulos \& Ivison 2002).

Another limitation  of the standard  method is encountered  in tracing
metal-poor  H$_2$  gas, which  becomes  more  severe  in enhanced  FUV
radiation  fields: the  reduced dust-shielding  and  CO self-shielding
allows  UV   photons  to  dissociate  CO  while   leaving  the  mostly
self-shielding  H$_2$  intact (Maloney  \&  Black  1988; Israel  1988,
1997).   This is now  corroborated by  observations of  the metal-poor
outer parts of typical spirals (Nakai \& Kuno 1995; Arimoto, Sofue, \&
Tsujimoto  1996),  globally  metal-poor  objects like  the  Magellanic
Clouds and Magellanic Irregulars (e.g.  Madden et al.  1997), and blue
compact  dwarf galaxies  (e.g.  Barone  et al.   2000).  Unfortunately
both types  of the aforementioned  bias can hinder efforts  to measure
H$_2$ gas mass  at high redshifts since high-$J$  $ ^{12}$CO lines are
those  used most  often,  and significant  amounts  of metal-poor  gas
reside in UV-intense environments (e.g. in Ly-break galaxies).

The CI J=1--0 line was one of the first Photodissociation Region (PDR)
lines to be detected from molecular clouds (Phillips \& Huggins 1981),
yet its H$_2$  tracing potential has gone unnoticed.   There are three
main reasons for  that namely, a) the low  atmospheric transmission at
its rest-frame frequency of  $\sim 492$ GHz precluded, until recently,
routine  observations  of  molecular  clouds in  this  transition,  b)
instrumentation constraints  in terms of receiver  sensitivity and the
ability to  map large areas fast (no  multi-beam receivers available),
and c) an early  theoretical prejudice of one-dimensional steady-state
PDR models  that pictured CI  distributed only in a  relatively narrow
CII/CI/CO transition zone on  the surface of FUV-illuminated molecular
clouds (e.g.  Tielens \& Hollenbach  1985a,b) and thus unable to trace
the bulk  of their  mass.  Interestingly this  prejudice was  at times
reinforced by  the instrumentation and  atmospheric restrictions which
allowed only limited mapping of  CI emission, which in turn was mainly
conducted in regions where theory predicted CI should~be.  These early
views  on the  CI  distribution  in PDRs  could  explain some  general
features of its  emission such as the surprising  robustness of the CI
J=1--0  line brightness  over  a  wide range  of  FUV intensities  and
metallicities  (e.g. Kaufman  1999).

This simple picture is now challenged by two large scale CI surveys of
the  Orion A  and B  molecular  clouds (Ikeda  et al.   2002) and  the
Galactic Center  (Ojha et al.  2001)  that find CI, $  ^{12}$CO, and $
^{13}$CO J=1--0  emission to be fully  co-extensive, their intensities
tightly  correlated, and  with a  surprisingly constant  $\rm N(CI)/N(
^{12}CO)\sim 0.1-0.2$ ratio over dramatic changes in ambient FUV field
and physical conditions.  These observations along with a considerable
body  of similar  past observational  evidence (see  Keene 1997  for a
review) point to a CI  distribution fully concomitant with that of CO.
The  present  work uses  time-varying  chemical  models under  typical
conditions found in Giant Molecular Clouds (GMCs) to further elucidate
this issue, and then examines the  potential of CI line emission as an
H$_2$ mass tracer.  We focus mainly on the following issues,

\noindent
1) The  current state  of the  art  of CI  observations revealing  its
distribution  throughout molecular clouds  (thus allowing  it to  be a
tracer of their mass), and  the role of non-equilibrium and/or dynamic
processes as well as cosmic rays under typical ISM conditions.

\noindent
2) The  expected CI  line  flux  densities over  a  range of  physical
conditions and  their relative  observational advantage (if  any) with
respect to the $ ^{12}$CO transitions.

\noindent
3) The uncertainties  of H$_2$  mass estimates from CI  emission with
 respect  to those associated  with the  standard method  employing CO
 lines, as well  as the observational prospects of  using the CI lines
 to trace molecular gas at high redshifts.

\noindent
Throughout  this  work, we  adopt a $\rm  q_{\circ  }=0.5$ cosmology
with $\rm \Omega_{\Lambda}=0$.

\section{The distribution of CI in GMCs}

In an early attempt to account for the mounting number of observations
finding  CI to be  more extended  than predicted  by the  models (e.g.
Keene et al.  1985) low density PDRs ($\rm n\sim 100$ cm$ ^{-3}$) have
been  advocated. There the  CI abundance  in the  CII/CI/CO transition
zone becomes  larger ($\rm [C]/[  ^{12}CO]\sim 1-100$) than  in denser
ones  ($\rm  n \geq  10^4\  cm^{-3}$)  because  a different,  CI-rich,
chemical  route is  favored, allowing  CI  to dominate  over a  larger
region  of the  cloud, $\rm  A_v\sim 3-7$  (Hollenbach,  Takahashi, \&
Tielens  1991).   More realistic  inhomogeneous  PDR  models showed  a
deeper FUV  penetration and a  more pervasive CI  emission ``coating''
the clumps making up the  GMCs (Meixner \& Tielens~1993; Spaans \& van
Dishoeck~1997).

The  aforementioned schemes  can produce  a {\it  spatial} correlation
between  CI  emission  and that  of  H$_2$  mass  tracers like  the  $
^{12}$CO,  $  ^{13}$CO J=2--1,  1--0  transitions  but cannot  readily
explain their tight {\it  intensity} correlation (Keene et al.  1997).
The latter is surprising because e.g.  a CI/$ ^{13}$CO intensity ratio
should depend on quantities like the surface-layer to total cloud mass
ratio, the different ambient  excitation conditions of the two species
(even  in clumped PDRs  CI and  $ ^{13}$CO  still reside  in different
volumes), all expected  to vary strongly within a  molecular cloud and
between  clouds. The  discovery  of significant  CI  emission deep  in
UV-protected sites of cold dark clouds in the Galaxy (e.g.  Oka et al.
2001), and  M 31  (Israel, Tilanus, \&  Baas 1998) makes  the standard
steady-state PDR  interpetation (with typical cosmic  ray fluxes) even
more difficult.   In the few cases  where CI J=2--1  is also detected,
the CI (2--1)/(1--0) ratio yields (in LTE) similar gas temperatures to
those  deduced solely  from the  $ ^{12}$CO  lines or  FIR/sub-mm dust
continuum in  either starburst  or quiescent environments  (Stutzki et
al.  1997 and references~therein).  Finally the striking similarity of
CI and  $ ^{13}$CO line profiles  in the latest large  scale survey by
Ikeda  et al.   (2002) suggests  a spatial  and  intensity correlation
continuing to  still smaller scales  than those probed by  the nominal
resolution of  the particular observations of $\sim  0.3$ pc (assuming
macroscopic motions  of much smaller,  radiatively-decoupled ``cells''
responsible for line-formation in  GMCs, see e.g.  Tauber 1996).  {\it
The  simplest  explanation  of  all the  aforementioned  observational
results is  that CI and  CO are fully  concomitant and trace  the same
H$_2$ gas mass.}

\subsection{Is the [C]/[CO] equilibrium reachable?}

 Non-equilibrium  chemical  processes  can  maintain  a  high  [CI]/[$
^{12}$CO] in cloud regions that have not attained chemical equilibrium
(e.g.  de  Boisanger \&  Chi\'eze 1991; Lee  et al.   1996; St\"orzer,
Stutzki, \& Sternberg 1997), or have done so under turbulent diffusive
``mixing'' of a  CI-rich cloud envelope with its  deeper regions (Xie,
Allen, \&  Langer 1995; Xie~1997).   The most recent  investigation of
the time dependance of the [C]/[$ ^{12}$CO] ratio and the resulting CI
line intensities by  Sto\"rzer et al.  examines the  effects of abrupt
changes of a strong local FUV field intensity ($\rm G_{\circ}\sim 10^5
$) caused by cloud-cloud shadowing over timescales of $\la 10^5$~yrs.

 In the  present work  we assume ambient  conditions expected  for the
bulk  of the  molecular gas  in GMCs  rather than  in their  few  O, B
star-illuminated  hot/dense spots.  These  conditions consists  of: a)
initially  atomic clouds (the  likely precursors  of GMCs),  b) softer
ambient FUV fields, and c) a larger range of gas densities.  Moreover,
processes  other than  cloud-cloud shadowing  govern the  evolution of
GMCs  and their ambient  conditions, with  timescales short  enough to
allow them  to maintain a  high [C]/[$ ^{12}$CO] ratio  throughout the
volume of a typical GMC during its lifetime.

In  the  case of  chemical  processes  the  H$_2$ formation  timescale
provides a good approximation for an overall equilibrium to be reached
in a  molecular cloud  that is initially  atomic (e.g.   Hollenbach \&
Tielens 1999). This timescale~is

\begin{equation}
\rm t_{ch}\sim (2\langle n \rangle R_f)^{-1}= 10^7 \left(\frac{T}{100 K}
\right)^{-1/2}\
 \left(\frac{\langle n \rangle}{50\rm cm^{-3}}\right)^{-1}\rm yrs,
\end{equation}

\noindent
where  $\rm \langle  n \rangle  $ is  the average  H density  and $\rm
R_f=3\times 10^{-18} (T/K)^{1/2}\ cm^3\ s^{-1}$ is the canonical H$_2$
formation rate on  dust grains (e.g.  Jura 1975).   For $\rm \langle n
\rangle \sim  (20-60)$ cm$ ^{-3}$ and $\rm  T\sim (70-120)$~K (typical
for the Cold Neutral  Medium HI out of which GMCs form),  it is $\rm t
_{ch}\sim  (1-3)\times  10^7$~yrs.  On  the  other  hand, the  dynamic
evolution timescale is

\begin{equation}
\rm t_{dyn}=\left(\frac{3\pi}{16G \langle \rho \rangle}\right)^{1/2} \sim
0.9 \times 10^7 \left(\frac{\langle n \rangle}{50 cm^{-3}}\right)^{-1/2} yrs,
\end{equation}

\noindent
(spherical cloud assumed). This is  comparable to $\rm t _{ch}$ and is
in  accordance with  the  well  known fact  that  timescales of  $\sim
(10^6-10^7)$~yrs characterize  a wide variety of  processes that fully
disrupt or otherwise drastically alter typical GMCs.  Some of the most
important ones are star formation with the disruptive effects of O, B,
star  clusters   (e.g.   Bash,  Green,  \&   Peters  1977),  turbulent
dissipation (Stone,  Ostriker, \& Gammie  1998; MacLow et  al.  1998),
and inter-cloud clump-clump collisions  (Blitz \& Shu 1980).  The fact
that  these  timescales are  shorter  or  comparable  to $\rm  t_{ch}$
suggests that molecular clouds may never reach chemical~equilibrium.

Driven by  the evolution  of continuously forming  stellar populations
throughout  a  galaxy, the  {\it  average}  FUV  field regulating  the
CII/CI/CO  transition  zone  in  stationary PDR  models  also  varies.
Recent work by Parravano, Hollenbach \& McKee (2003) has shown

\begin{equation}
\rm t _{FUV}=k \lambda ^{-1} _{SFR}\left(\frac{U}{U_{\circ}}\right)^{3/2}
 10^6 \ yrs,
\end{equation}

\noindent 
to be  the typical time needed  for a significant  perturbation of the
 FUV energy density  to occur, where $\rm k=1-2$, U  is the FUV energy
 density,  $\rm U_{\circ}=10^{-17} ergs\  cm^{-3} \AA^{-1}$,  and $\rm
 \lambda _{SFR}$  is the star  formation rate normalized to  its Solar
 neighborhood value.  For e.g.   $\rm U=10\times U_{\circ}$ it is $\rm
 t _{FUV}\sim  5\times 10^7\ yrs$,  comparable to $\rm  t_{ch}$
 even in quiescent environments ($\rm \lambda _{SFR}=1$).

In order to  track the evolution of the  [C]/[$ ^{12}$CO] ratio during
the aforementioned timescales we present results from our homogeneous,
time-dependent  PDR models  in  Figure  1.  The  chemistry  of gas  is
represented  over a semi-infinite  slab, taking  into account  all the
heating and cooling processes (see Papadopoulos, Thi, \& Viti 2002 and
references therein).  We have also included the latest advancements in
collisional/reaction  rates and  chemical networks  which  now include
most charge-exchange reactions.  The latter are particularly important
in  the proper  estimate of  the  [C]/[$ ^{12}$CO]  values in  heavily
shielded, cosmic ray-dominated environments with various metallicities
(sections 2.2, 2.3).

\begin{figure}
\resizebox{\hsize}{!}{\includegraphics[angle=90]{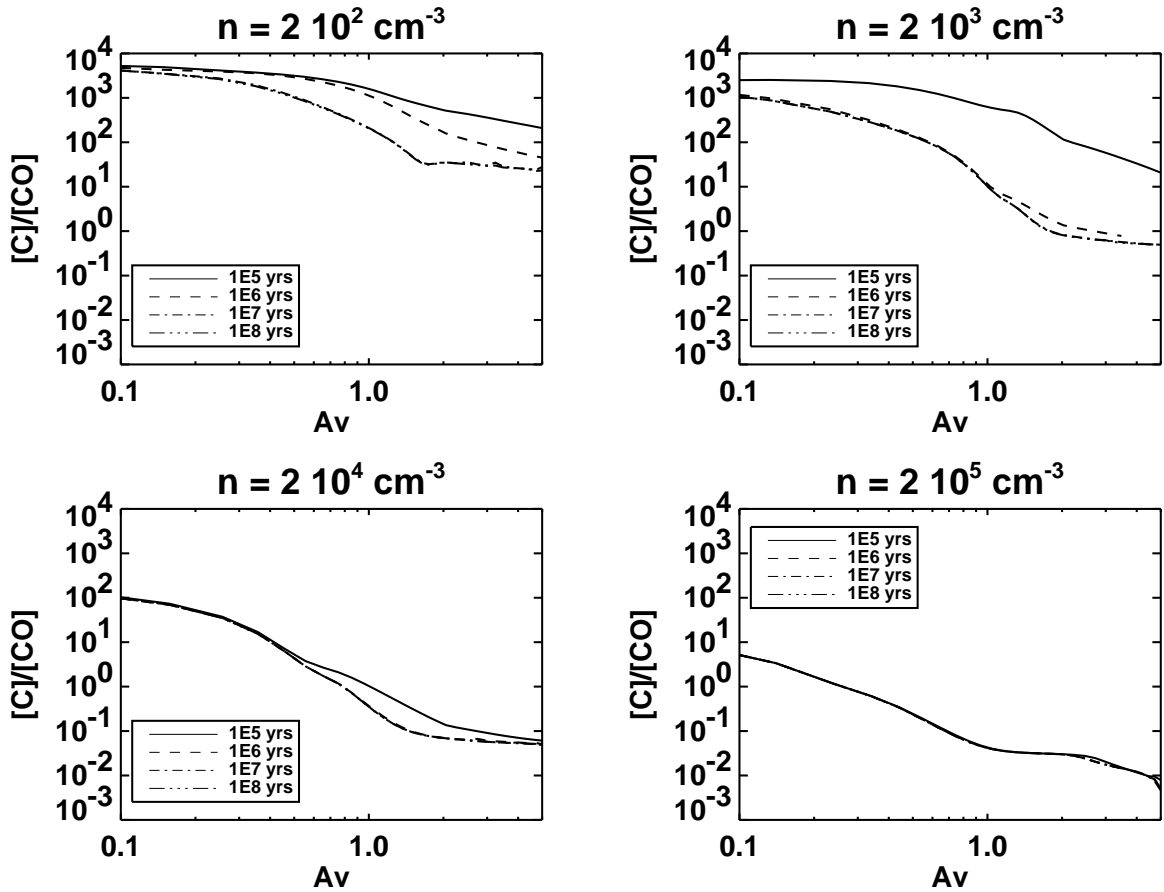}}
\caption{Results  from a  time-dependent PDR  code (see  text)  for an
initially atomic cloud with one-sided illumination by a unidirectional
FUV field  and densities shown at  the top of each  frame. The assumed
metallicity is  solar (Z=1) and  the incident radiation field  is $\rm
G_{\circ}=1/2$, (corresponding  to one-sided illumination  by a Habing
field). The well-known [C]/[$ ^{12}$CO] enhancement in low density gas
with respect to denser gas is clearly discernible }
\end{figure}

 The  diagrams  in  Figure   1  confirm  that  the  [CI]/[$  ^{12}$CO]
equilibrium  timescale   and  $\rm  t  _{dyn}$   (Equation  2)  remain
comparable  throughout  the  density  range  examined.   Hence,  since
self-gravity is  the plausible  main ``driver'' behind  many processes
altering  the physical  conditions  of a  GMC  (e.g.  star  formation;
Elmegreen  2002), it  can  be argued  that  [CI]/[$ ^{12}$CO]  remains
significant,  never   settling  to  its   (small)  equilibrium  values
throughout its lifetime.  In more active star forming environments all
processes that disrupt GMCs  or alter their ambient conditions operate
on  still shorter  timescales.  The  effect of  higher  star formation
rates  is easily  highlighted by  Equation  3 since  for $\rm  \lambda
_{SFR}>1$, $\rm t_{FUV}$ simply becomes proportionally~shorter.

 In the absence of a  generally accepted GMC formation and destruction
theory  the most  general way  to view  the effects  of  elevated star
formation on chemical equilibrium  is by assuming that star formation
powers the cycling of interstellar gas between the Warm, Cold HI (WNM,
CNM), and  the molecular gas  phases.  Then the gas  condensation rate
$\rm  \omega (n_H)=  d(log  n_H)/dt=\phi _{SFR}/g(n_H)$,  where $  \rm
g(n_H)=\partial  M/\partial   (logn_H)$  is  the   mass  fraction  per
logarithmic  density  interval (Lioure  \&  Chi\'eze 1990).   Clearly,
irrespective of the particular physical processes defining $\rm \omega
(n_H)$, a higher  $\rm \phi _{SFR}$ induces a  faster evolution of any
given gas  ``parcel'' across the  entire mass-density spectrum  of the
neutral ISM, leaving less time for it to reach chemical~equilibrium at
any one phase.

\subsection{The role of turbulence}

A relatively recent observational  result that is particularly hard to
reconcile with  steady-state models was  obtained by the  extensive CI
J=1--0 and  $ ^{12}$CO J=4--3  imaging of the Carina  molecular clouds
where  a close spatial  association of  their emission  throughout the
cloud was found, suggesting that  the same gas emits both lines (Zhang
et  al.  2001).   Low density  PDRs ($\sim  10^2$ cm$^{-3}$)  would be
CI-rich  and bright  in CI  J=1--0 but  not in  the $  ^{12}$CO J=4--3
transition that has $\rm  n_{43}\sim 2\times 10^4\rm \ cm^{-3}$, while
in the  bi-stability scenario (section  2.2) a locally much  higher CR
flux (so that $\rm n_{(CR)}  \ga 10^4\rm\ cm^{-3}$) would be needed to
explain such~association.

Turbulent diffusion can smooth any initial [CI]/[$ ^{12}$CO] abundance
  gradient  so that it  becomes almost  constant throughout  a typical
  cloud, and in the Carinae  molecular complex the clouds are found to
  be warmer  than typical  ones in the  disk, suggestive  of turbulent
  velocity  field dissipation  (Zhang et  al. 2001).   The  effects of
  turbulent diffusion  on elemental  abundances have been  outlined by
  Xie, Allen, \&  Langer (1995) (see also Xie 1997  for a review), and
  here we will only point out the effects in the kinematically violent
  environments found in e.g.  galactic centers and merger systems.  In
  such regions  molecular clouds have very  high velocity dispersions,
  typically  $\rm \sigma  _{V}\sim 50\  km\ s^{-1}$  for  the Galactic
  Center (e.g.   Oka et  al.  1996),  that can reach  up to  $\rm \sim
  (50-100)\ km\  s^{-1}$ in the  extreme ULIRG merger  systems (albeit
  model-dependent,  e.g.  Downes  \&  Solomon 1998).   These are  much
  higher than  $\sim \rm  (1-5)\ km\  s^{-1} $ found  for GMCs  in the
  kinematically quiescent  spiral disks.   These large line  widths in
  kinematically   violent  environment   correspond  to   a  turbulent
  diffusion   coefficient   $\rm   K=   \langle  V_t   L   \rangle\sim
  (10^{24}-10^{25})\ cm^2\ s^{-1}$,  (assuming $\rm \sigma _V\sim V_t$
  and an invariant characteristic length scale L).  This is 1-2 orders
  of magnitude larger than in kinematically quiescent environments and
  can  easily  yield a  constant  [C]/[$  ^{12}$CO]$\sim 0.1-1$  ratio
  throughout the  cloud (Xie et al.  1995).   The diffusion timescales
  are  short, e.g.   for  quiescent GMCs  $\rm  t_{diff} \sim  3\times
  10^6$~yrs,  which is shortened  to $\rm  \sim 3\times  (10^4-10^5) \
  yrs$  in  dynamically  active  environments ($\rm  t_{diff}  \propto
  1/K$).

Such effects, and  a potentially large cosmic ray  flux (section 2.3),
can easily account for  the significant [CI]/[$ ^{12}$CO] ratios found
in dark and  dense clouds, shielded from the  ambient FUV radiation by
very  large extinctions (e.g.   Ikeda et  al.  2002),  as well  as the
similarity  of CI  J=1--0 and  HCO$^+$  J=1--0, 3--2,  CS J=2--1  line
profiles (Stark et al.  1997; Johansson et al.  1994).  The latter are
particularly  puzzling since  in  shielded environments  HCO$^+$ is  a
product of dense, dark cloud  chemistry, and its J=3--2 transition has
$\rm  n_{cr}\sim 10^6$~cm$ ^{-3}$  which, along  with CS  J=2--1 ($\rm
n_{cr}\sim 10^5\  cm^{-3}$), imply very  dense gas for  which standard
PDR  models  predict $  ^{12}$CO  to  be  the dominant  carbon-bearing
species~($\rm [CI]/[ ^{12}CO]<10^{-3}$).   Turbulence has the strength
and  the  time  to  redistribute  any CI-rich,  PDR-origin  gas  phase
initially  located  in a  cloud  envelope  throughout  its denser  and
UV-shielded interior within~its~lifetime.

\subsection{Processes controlling [CI]/[$ ^{12}$CO]: the role
of cosmic rays}

The previous  brief exposition of time-dependent  chemical and dynamic
 phenomena suggests  that {\it  except in few  of their  densest, most
 UV-protected  and  kinematically quiescent  regions,  GMCs may  never
 attain chemical equilibrium while turbulence is capable of ``mixing''
 the resulting CI-rich gas phase throughout their volume}, a situation
 accentuated in star forming and/or dynamically active environments.

In star forming environments the  expected larger cosmic ray flux will
 also raise the [CI]/[$ ^{12}$CO]  ratio deep inside a cloud's volume,
 and several  observations have  indeed found a  systematically larger
 [CI]/[$ ^{12}$CO]  ratio in starburst  nuclei (e.g.  Harrison  et al.
 1995; Israel \& Baas 2002).  In  the starburst M 82, bright CI J=1--0
 emission with  $\rm N(CI)/N(  ^{12}CO)\sim~0.5$ is measured  over the
 bulk of  its H$_2$  gas (Schilke  et al.  1993;  White et  al.  1994)
 while  the  CR  flux  in  its  central  regions  is  $\rm  F_{CR}\sim
 (170-500)\times F^{(Gal)} _{CR}$ (Suchkov, Allen, \& Heckman~1993).

In  Figure 2  the effect  of $\rm  F_{CR}$ on  the CI  distribution is
 presented,   and  the   CR-induced   [CI]/[$  ^{12}$CO]   enhancement
 throughout a cloud becomes apparent (see also Flower et al.  1994 for
 early  such work).  Indeed  in the  dark cloud  conditions prevailing
 deep into the  inner regions of GMCs a pure PDR  origin of CI becomes
 unattainable but  cosmic rays, unlike  ultraviolet photons, penetrate
 deep into clouds  and at high $\rm A_v$  control both thermal balance
 and  chemistry.  Such  a  chemistry can  still  yield a  steady-state
 CI-rich gas  phase, provided $\rm n\la  n_{(CR)}$.  This ``turnover''
 density increases  for higher $\rm F_{CR}$ and  marks a CR-controlled
 bi-stability  of the chemical  network set  by the  abrupt transition
 between two possible ISM  ionization states (des For\^ets, Roueff, \&
 Flower 1992; Le Bourlot, des  For\^ets, \& Roueff 1993; Flower et al.
 1994).   For  $\rm n\la  n_{(CR)}$  the  gas  has a  high  ionization
 fraction  and thus a  high $\rm  H^+ _3$  dissociative recombination,
 allowing H$^+$ to be the dominant  ion and initiate a set of chemical
 reactions that  yield $\rm [CI]/[ ^{12}CO]\sim 0.1-0.2$.   For $\rm n
 >n_{(CR)} $, $\rm  H^+ _3$ becomes dominant and  a different chemical
 network yields a much lower $\rm [C]/[ ^{12}CO]\la 10^{-3}$.

 For  the Galaxy  it  is  $\rm n_{(CR)}  \sim  5\times 10^3\  cm^{-3}$
(Schilke et  al.  1993 and  references therein), encompassing  a large
fraction  of   the  molecular  gas  in~GMCs.    For  totally  shielded
environments ($\rm G_{\circ}=0$)  and $\rm F_{CR}=F^{(Gal)} _{CR}$ our
models yield $\rm [C]/[ ^{12}CO]\ga  3$ for $\rm n\leq 10^3\ cm^{-3}$,
and  $\rm  [C]/[  ^{12}CO]\leq  0.02$  for $\rm  n\geq  5\times  10^3\
cm^{-3}$.    A  ten-fold   increase   in  $\rm   F_{CR}$  boosts   the
aforementioned densities by an  order of magnitude, for similar [C]/[$
^{12}$CO]  values.  These  values are  in good  accordance  with those
reported  in the  literature even  though controversies  remain  and a
major benchmarking effort  of the various PDR codes  is under way.  In
an important recent development the canonical value of $\rm F_{CR}$ in
quiescent  diffuse clouds  in the  Solar neighborhood  may have  to be
revised upwards  by a factor of  $\sim 40$ (McCall et  al.  2003) thus
opening up  the possibility of a  mainly CR-origin of  the CI emission
observed to be widely distributed in the quiescent molecular clouds in
the disk of the Galaxy.

\begin{figure}
\resizebox{\hsize}{!}{\includegraphics[angle=90]{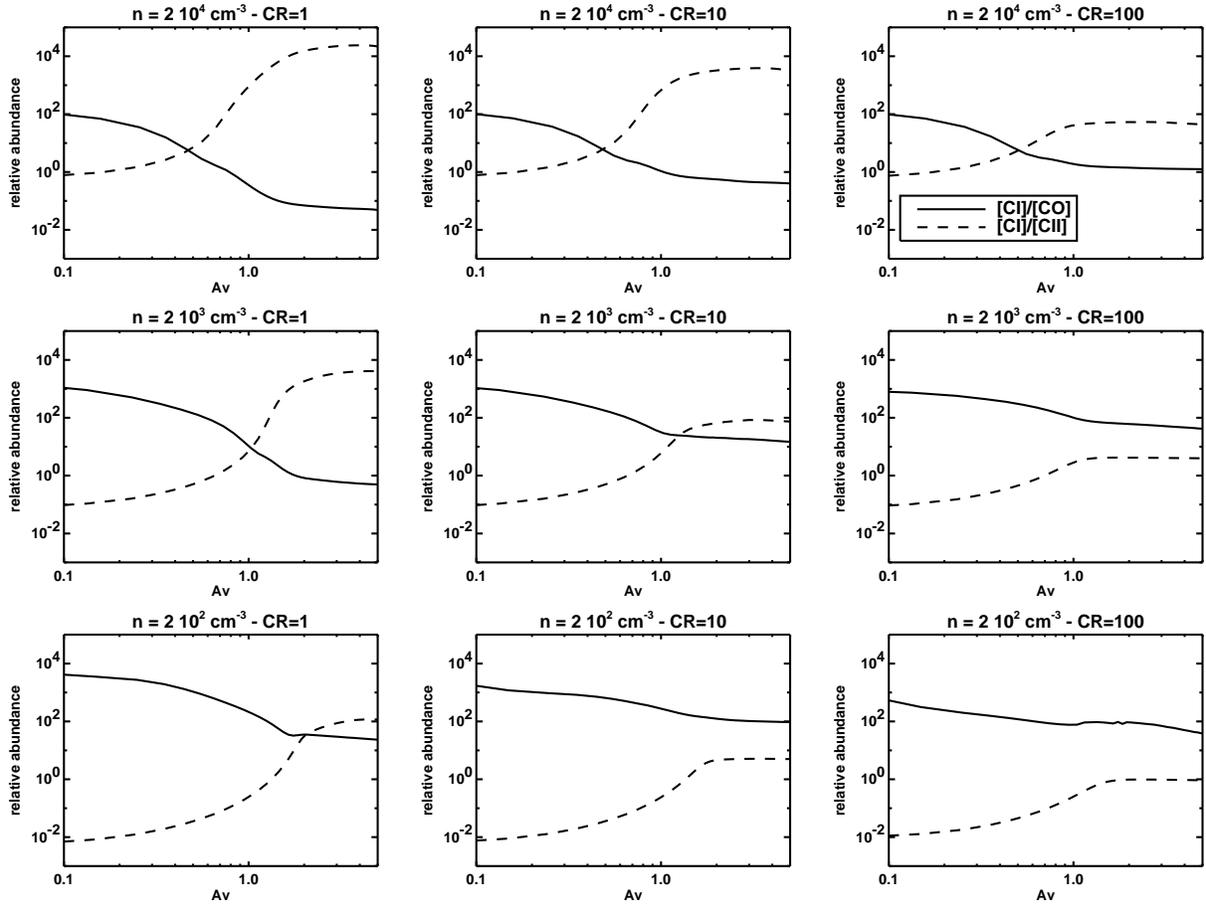}}
\caption{Abundance  distributions at $\rm  t=10^7$ yrs,  for identical
 initial conditions  as in Figure 1.   The values of  $\rm F_{CR}$ are
 indicated  in  the panels,  normalized  to  the  Galactic value.  The
 ambient  $\rm   G_{\circ}=1/2$,  corresponding  to   one-sided  cloud
 illumination by a Habing field.   In most cases [CI]/[CII]$\ga 1$ for
 $\rm A_v\ga  1$. The density  range chosen characterizes the  bulk of
 H$_2$ in typical~GMCs.}
\end{figure}

 In Figure 3 we explore the  same effects but now scale {\it both} the
CR flux and the FUV ambient radiation field, since their {\it average}
values  are expected  to  rise  in tandem  in  realistic star  forming
environments.   In this  case  enhanced CII  production  at large  FUV
intensities limits the  cloud volume where CI is  dominant, but except
for diffuse gas  irradiated by strong fields, CI  remains the dominant
form of carbon for $\rm A_v\ga 2$.

\begin{figure}
\resizebox{\hsize}{!}{\includegraphics[angle=90]{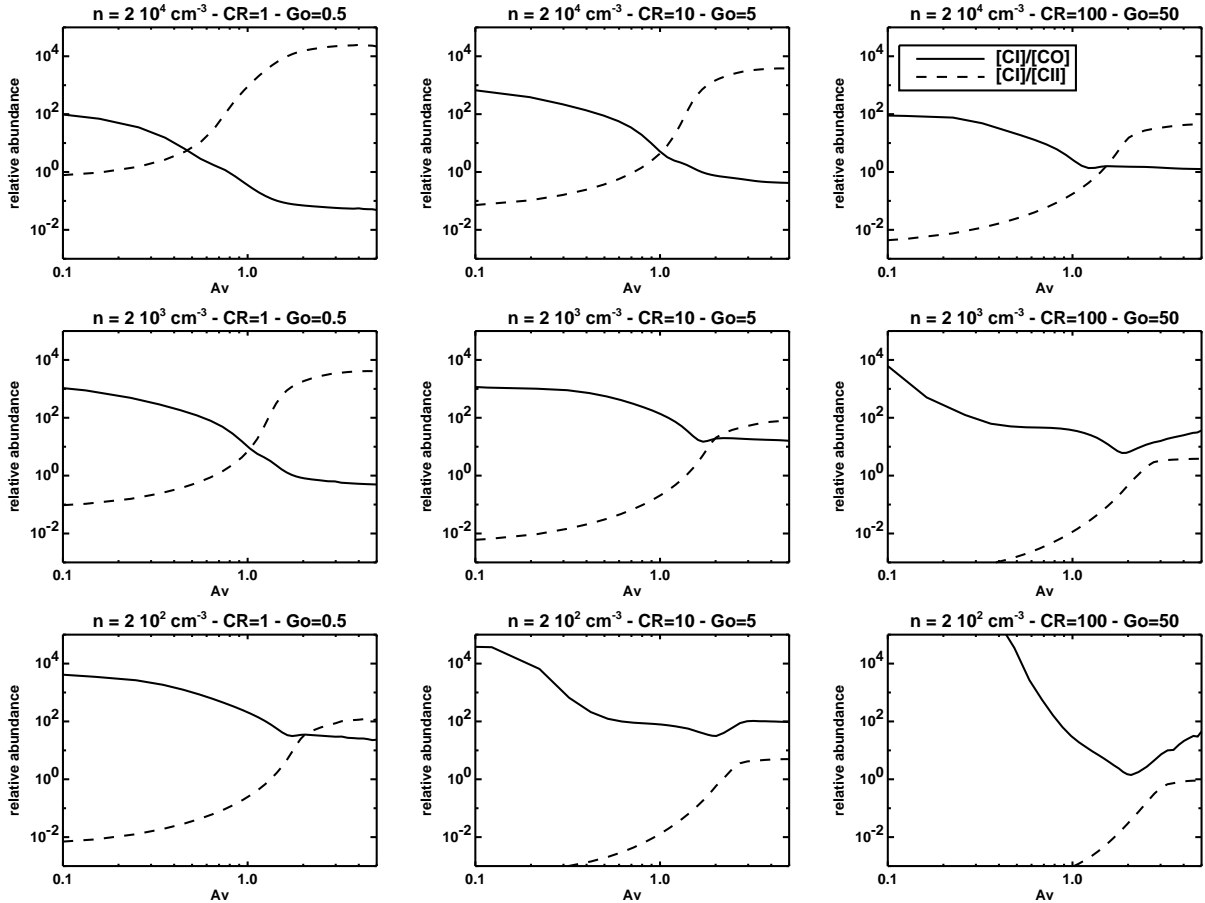}}
\caption{Abundance  distributions at  $\rm t=10^7$  yrs,  assuming the
same initial  conditions as in Figure  1.  The values  of $\rm F_{CR}$
are indicated  in the  panels, normalized to  the Galactic  value. The
ambient  $\rm  G_{\circ}$  is   now  scaled  proportionally  but  also
multiplied by 1/2 corresponding to one-sided cloud illumination.}
\end{figure}

\subsection{CI distribution in metal-poor environments}

The  expected evolution  of metallicity  in galaxies  with  an overall
trend of  lower values at earlier  epochs is now  verified by numerous
observations (e.g.   Mehlert et  al.  2002).  A  trend of  higher star
formation  rates in  galaxies  at higher  redshifts  is also  deduced,
although the details are far from clear (e.g.  Ellis 2001; Somerville,
Primack, \&  Faber 2001).  It  is therefore anticipated  that galaxies
with low metallicity  and elevated star formation rates  (and thus FUV
and CR  flux), will  be common at  high redshifts.   Ly-break galaxies
found at $\rm z\sim 3$ with moderate star formation rates of $\rm \sim
(5-10)\rm\ M_{\odot  } yr^{-1}$ (Steidel et al.   1996), and sub-solar
metallicities (e.g.  Pettini et al.  2000) are typical such objects.

In low-metallicity  environments with pervasive  FUV radiation powered
by star  formation, the  enhanced CO photodissociation  diminishes its
spatial  extent  per  molecular   cloud  and  thus  its  H$_2$-tracing
capability  (Pak et  al.  1998;  Bolatto, Jackson,  \&  Ingalls 1999).
Observational work using the AST/RO facility in the South Pole Station
confirms these  expectations by finding enhanced  CI/$ ^{12}$CO J=1--0
line ratios  in the LMC  and SMC (Stark  et al.  1997; Bolatto  et al.
2000a).   These  data, albeit  sparse  and  with  a large  dispersion,
suggest  $\rm   I_{CI}/I_{CO}\propto  Z^{-1/2}$,  supported   also  by
observations  in  the  low-metallicity  star forming  irregular  IC~10
(Bolatto et al.  2000b and references therein).

\begin{figure}
\resizebox{\hsize}{!}{\includegraphics[angle=90]{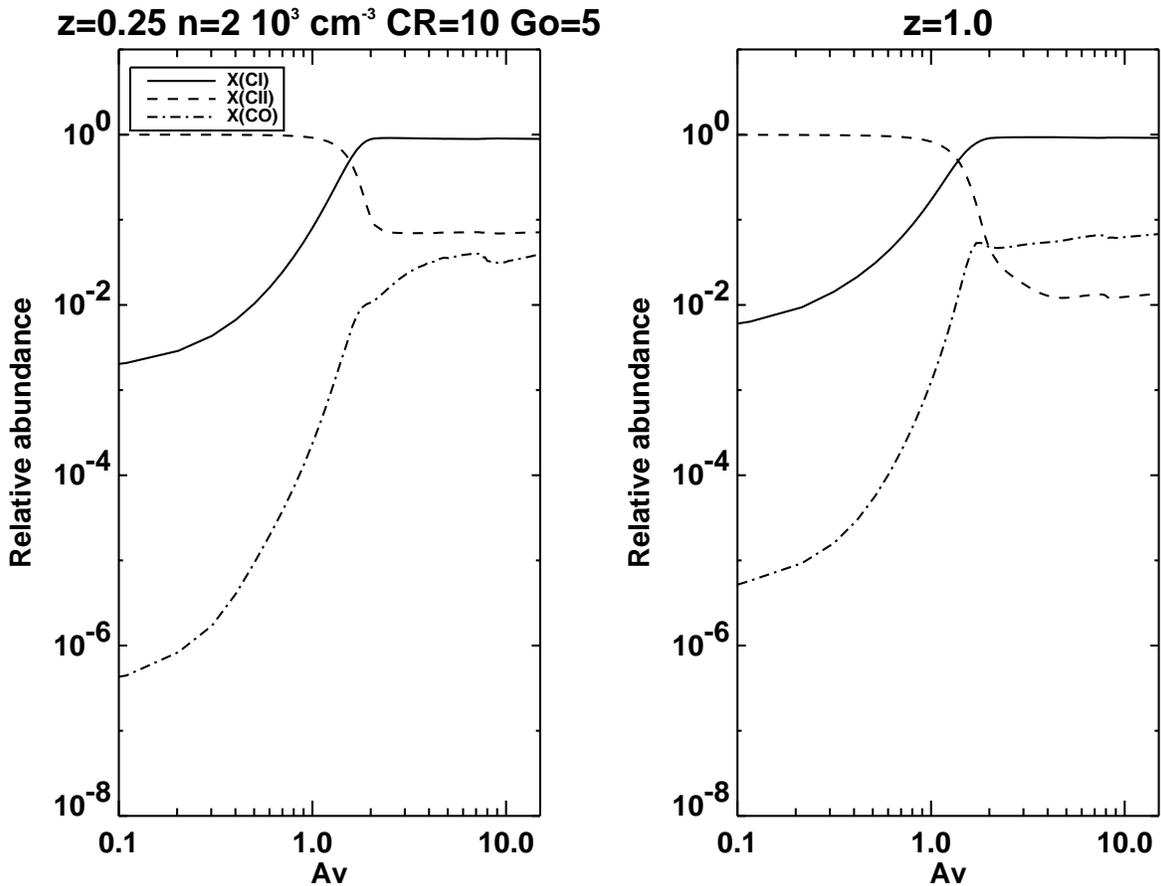}}
\caption{Carbon  species abundance distributions  at $\rm  t=10^7$ yrs
{\it  relative to  the total  carbon  content}. The  cloud is  assumed
initially atomic,  and $\rm F_{CR}$, $\rm G_{\circ}$  correspond to an
underlying  star formation~rate  of $\sim  10\rm  M_{\odot}\ yr^{-1}$,
typical of Ly-break galaxies.   The sub-solar metallicity of Z=0.25 is
also typical for Ly-break galaxies (Pettini et al. 2000).}
\end{figure}

In Figure 4  the effects of such conditions on  the CI distribution in
solar and  sub-solar metallicity  environments are contrasted,  and it
can be seen  that CI remains the dominant form of  carbon for the bulk
of the cloud ($\rm A_v\geq 2$) even in the metal-poor case.  {\it Thus
in   the   metal-poor    and   modestly   FUV-enhanced   environments,
characterizing  many galaxies  at high  redshifts, CI  remains  a good
tracer of molecular gas mass.}

A large  change is seen however  in the distribution of  CII that now,
unlike in the  metal-rich case, it becomes more  abundant than CO even
deep into the cloud (yet still less than CI).  Apart from the enhanced
FUV photoinization  of CI  at low $\rm  A_v$'s, caused by  the reduced
dust  shielding, low metallicity  further enhances  CII at  the deeper
parts of such clouds at the expense of CO because the main route of CO
destruction at high $\rm A_v$ is via the reaction path: $\rm CO + He^+
\rightarrow C^+ + O + He.$ Once formed, ionized carbon will capture an
electron from  a metal atom in  the gas phase but  at low metallicities
the charge exchange reactions involving the metallic atoms and ions are
less efficient  in converting C$^+$ into  C while He$^+$  is formed in
large abundance by the larger  cosmic rays flux and thus a significant
part of  the carbon  will be  in the form  of CII  even at  large $\rm
A_v$'s.

\newpage

\section{The intensity  of the CI lines}

The   critical   densities   of    the   fine   structure   CI   lines
$^{3}P_{1}\rightarrow$$^{3}P_{0}$  ($\nu   _{10}=492.160$~GHz)  and  $
^{3}P_2\rightarrow$$^{3}P_1$   ($\nu  _{21}=809.343$  GHz)   are  $\rm
n_{10}\sim  500$ cm$ ^{-3}$  and $\rm  n_{21}\sim 10^{3}$  cm$ ^{-3}$,
comparable to  those of  the widely observed  $ ^{12}$CO  J=1--0, 2--1
transitions, while their upper energy levels are $\rm E_1/k = 23.6$~K,
and $\rm E_{2}/k=62.4$~K.  The  velocity-integrated flux density of a
line is

\begin{equation}
\rm \int _{\Delta v}  S_{x} dV = \frac{2 k \nu ^2 _{x}}{c^2} 
\left(\frac{1+z}{D^2 _L}\right) L_{x},
\end{equation}
  
\noindent
where $\rm D_{\rm  L}= 2 c H_{\circ } ^{-1}  (1+z -\sqrt{1+z})$ is the
luminosity   distance  for   a  redshift   z,  $\rm   L_{x}$   is  the
area/velocity-integrated  line brightness  temperature (in  K  km sec$
^{-1}$ pc$ ^2$), and $\rm  \nu _{x}$ its rest-frame frequency.  Hence,
for any of  the CI J=1--0, 2--1 and the  $ ^{12}$CO J+1$\rightarrow $J
lines the velocity-integrated flux density ratio is

\begin{equation}
\rm \frac{\int _{\Delta v}  S_{CI} dV}{\int _{\Delta v}  S_{CO} dV}=\left(\frac{\nu _{ci}}{
\nu_{co}}\right)^2 \frac{R_{CI}}{R_{J+1,J}},
\end{equation}

\noindent
where      $\rm       R_{CI}=      \langle      T_b(CI)\rangle/\langle
T_b(CO_{1-0})\rangle$,      and      $\rm      R_{J+1,J}      =\langle
T_b(J+1,J)\rangle/\langle   T_b(1,0)\rangle$  for   $   ^{12}$CO.  The
brightness temperature averages are over velocity and emitting area.

An unresolved (or barely resolved) object, whose various ISM lines are
accessible by  a multiplicity of  receivers, will have an  observed CI
flux density  larger by the  $(\nu _{\rm ci}/\nu _{\rm  co})^2$ factor
relative to  that of lower frequency  CO transitions.  Interferometric
scaled-array observations  `synthesizing' the same  beamsize at widely
different  frequencies  while   utilizing  the  same  collecting  area
maintain  this advantage  in the  case  of resolved  objects.  A  flux
density  advantage  of  a  CI  line  may remain  even  when  $\rm  \nu
_{ci}/\nu_{co}\la  1$, simply  because  CI lines  are  much easier  to
excite and thus can maintain  their intensities intact for diffuse and
cooler gas while the $\rm R_{J+1,J}$ ratios diminish (see also Kaufman
1999).   For example  the  $  ^{12}$CO J=4--3  and  J=7--6 lines  have
critical   densities  $\rm  n_{43}=2\times   10^4  \   cm^{-3}$,  $\rm
n_{76}=3\times  10^6\  cm^{-3}$  and  $\rm  E_4/k\sim 55  \  K$,  $\rm
E_7/k\sim 155\  K$, all much higher than  the corresponding quantities
of the CI J=1--0, 2--1 lines emitted at similar~frequencies.

Thus for any {\it detected $ ^{12}$CO J+1$\rightarrow $J (J+1$\geq $3)
line,  bright  CI  emission  may  also be  present,  particularly  for
sub-thermally  excited, and/or cooler  gas (low  $\rm R  _{J+1, J}$)}.
This  is of  particular importance  for  the mounting  number of  high
redshift (z$\ga  $2) CO  detections, where only  the $\rm J+1\geq  3 $
lines are  usually observed while for  such redshifts the  CI lines no
longer undergo large atmospheric absorption.

  In  Table 1  we summarize  the typical  brightness  temperature line
ratios  measured in  various ISM  environments in  the  local Universe
gleaned from an extensive search in the literature.  Large atmospheric
absorption   has   prevented  any   large   surveys   of  $   ^{12}$CO
J+1$\rightarrow$J,  J+1$>$3  or  $   ^{13}$CO  J=3--2,  so  we  use  a
single-phase Large Velocity Gradient (LVG) code, constrained by the CO
line  ratios  available  in  each  case,  to  estimate  expected  flux
densities  for  J+1$>$3.   The  results  are then  used  to  calculate
detection advantage  ratios $\rm D_a$, defined  as the signal-to-noise
(S/N)  ratio  normalized  by  that  of $  ^{12}$CO  J=1--0  for  equal
integration time  and velocity  resolution.  A general  expression for
the system sensitivity~is

\begin{equation}
\rm \delta S_{rms}=\left(\tau \Delta \nu \right)^{-1/2} \Delta S_{sys}=
\left[\frac{c }{\nu \tau _{int} \Delta V}\right]^{1/2} \Delta S_{sys}(\nu ),
\end{equation}

\noindent
where $\rm  \nu =(1+z)^{-1}\nu  _x$ is the  observed frequency  of the
line, $\rm  \tau _{int} $ is  the integration time, $\Delta  \rm V$ is
the velocity resolution, and  $\rm \Delta S_{sys}(\nu )$ describes the
overall  system   sensitivity  and  incorporates   instrumental  (e.g.
collecting    area,     receiver    temperatures)    as     well    as
atmosphere-dependent  factors (e.g.  absorption,  signal decorrelation
factors in case  of interferometers).  From Equations 4  and 6 the S/N
can be expressed as

\begin{equation}
\rm \frac{\langle S_{\nu }\rangle _{\Delta V}}{\delta S_{rms}}=
2k \left(\frac{\nu}{c}\right)^{5/2} \langle T_{b} \rangle 
\left[\frac{(1+z)^3}{D^2 _L}\right] \frac{(\tau _{int} \Delta V)^{1/2}}{\Delta
S_{sys}(\nu )},
\end{equation}

\noindent
where $\rm  \langle T_b \rangle  $ is the  velocity/area-averaged line
brightness  temperature.  For  frequencies  of $\sim  30-250$ GHz  and
future modern instruments like ALMA, located in dry sites, $\rm \Delta
S_{sys}(\nu   )/\nu  ^{1/2}   $  changes   $\la   50\%  $\footnote{See
http://www.alma.nrao.edu/info/sensitivities,  Table 1}.   Thus setting
$\rm \Delta  S_{sys}(\nu )\propto \nu  ^{1/2}$, and for a  common $\rm
\tau _{int}$ and $\rm \Delta V$,

\begin{equation}
\rm D_a = \frac{(S/N)}{(S/N)_{CO(1-0)}}=(\nu _x/\nu _{10})^{2} R
\end{equation}

\noindent
  expresses  the  detection  advantage  of  any  spectral  line  with
rest-frame frequency  $\rm \nu _x$,  normalized to $  ^{12}$CO J=1--0,
with  R being  their  brightness temperature  ratio.   This assumes  a
bandwidth fully  covering both lines  under comparison, which  will be
true for planned mm/sub-mm telescopes.

The $  \rm D_a$ values for  the $ ^{12}$CO  and CI lines are  shown in
Figure~5  where it is  obvious that  $ ^{12}$CO  J=4--3 marks  a broad
excitation  ``turnover'' beyond  which $  \rm D_a$($  ^{12}$CO) starts
declining.  This is  not surprising since for most of  the gas in GMCs
$\rm T_{kin}\la 50$~K, and  $\rm n(H_2)\la 10^4$~cm$ ^{-3}$, rendering
the  $  ^{12}$CO  J+1$\rightarrow$J,  J$\geq $~3  lines  sub-thermally
excited  and underluminous.  The  wide range  of $\rm  D_a$(CI) values
reflects  the considerable range  of $\rm  R_{CI}$ ratios  observed in
various environments, which is not  simply correlated to that of the $
^{12}$CO line  ratios, e.g.  a  large $\rm R_{J+1,J}$  not necessarily
mean  also a  large $\rm  R_{CI}$.   The CI  J=1--0 line  has a  clear
advantage over  $ ^{12}$CO  transitions in quiescent  environments and
maintains a sizable one in starbursts.  The CI $\rm J=2-1$ is the only
line in the  molecular ISM that maintains high $\rm  D_a$ at such high
frequencies  under  most  conditions,  a  conclusion  subject  to  the
uncertainty of  the assumed $\rm  CI (2-1)/(1-0)$ ratio which  has not
been  widely  measured.  For  metal-poor  starbursts  Figure~5 can  be
somewhat misleading  since in such environments CII  may dominate over
the bulk  of the molecular  gas mass with  both the $ ^{12}$CO  and CI
lines being rather weak.  Nevertheless in the case of a {\it detected}
$ ^{12}$CO line  in such an environment, CI  line emission is expected
to be bright as~well.

\begin{figure}
\resizebox{\hsize}{!}{\includegraphics[angle=90]{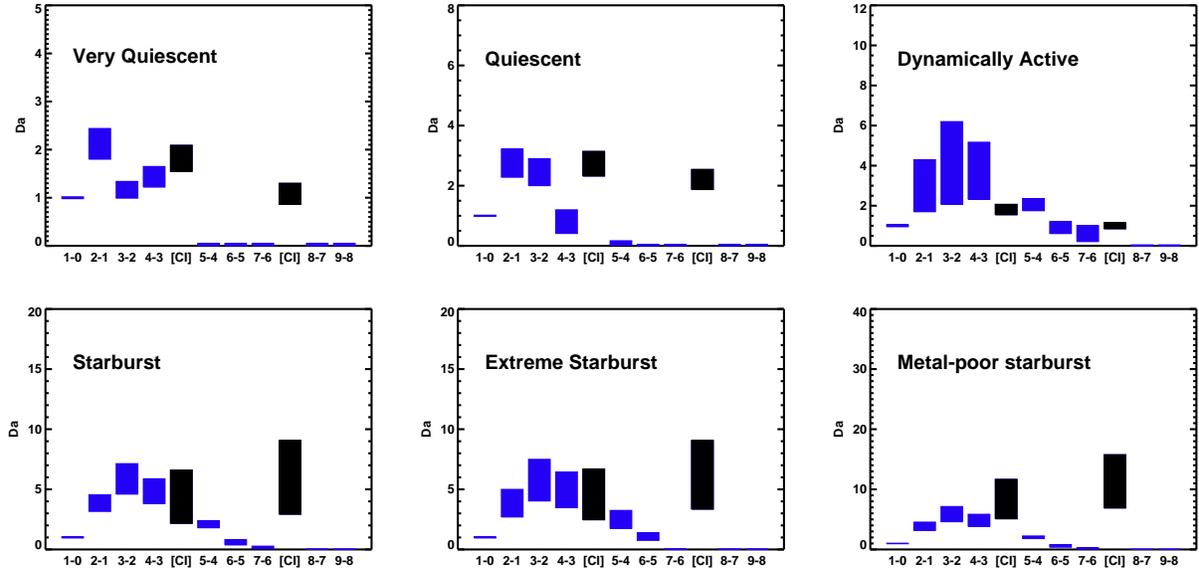}}
\caption{The values  of $\rm D_a$  for the ISM environments  listed in
Table~1 (see text).   A dispersion of $\sim 15\%$  was assumed for all
ratios  (typical error  of the  mean value  reported by  surveys). For
dynamically  active and  extreme starburst  environments we  adopt the
much larger range  of the CO line ratios found  around the mean values
reported in Table 1. For  metal-poor staburst we set Z=0.25, and adopt
the  scaling  $\rm   R_{CI}(Z)=Z^{-1/2}  R_{CI}(1)$  (Bolatto  et  al.
2000b), all other ratios are assumed the same as in starbursts.}
\end{figure}

It must be  emphasized that $\rm D_a$ values  translate to a practical
observational advantage  provided the relevant lines do  not fall near
any of  the (many) atmospheric absorption windows  within the $30-250$
GHz frequency  range, and that the  set of $ ^{12}$CO,  CI lines under
comparison is  accessible within this frequency  interval.  The latter
implies redshifts  of $\rm z_{1-0}\ga 1$ and  $\rm z_{2-1}\ga~2.25$ so
that the CI J=1--0, 2--1 lines fall within the aforementioned range.

 A  {\it  single}  receiver  observing  various  molecular  ISM  lines
redshifted within  its tuning  range from several  objects distributed
over  a   redshift  range   is  a  case   more  relevant   to  current
instrumentation capabilities (limited  set of receivers per telescope)
and  blank field  spectral line  surveys to  be conducted  with future
instruments  per observing  band  (e.g.  Blain  et  al.  2000).   Then
Equation~8  is  no longer  relevant  and one  has  to  simply look  at
Equation 7 and the expression  in the brackets which changes $\la 3\%$
for $\rm  z \sim  1-10$.  This  allows $ ^{12}$CO,  CI lines  to trace
similar amounts of gas mass over a progressively larger redshift range
as  long  as  sufficiently  excited lines  with  progressively  larger
rest-frame  frequencies   become  accessible.   This   is  simply  the
spectral-line equivalent of the well-known favorable K-correction that
keeps the FIR/sub-mm dust continuum of galaxies undiminished (and thus
detectable) over  a large  range of redshifts.   For $  ^{12}$CO lines
this is beautifully demonstrated by $ ^{12}$CO J=$1-0$ observations of
nearby QSOs that  measure H$_2$ masses similar to  those found in high
redshift QSOs and radio  galaxies detected through higher-J $ ^{12}$CO
lines (Evans et al.  2001).

Studies examining the luminosity of  molecular and atomic ISM lines in
the context of emission-line searches at high redshifts (e.g. Blain et
al.  2000) do not explore the H$_2$-tracing potential of the CI lines.
Moreover,  by making  favorable assumptions  about the  gas excitation
they  present optimistic predictions  of flux  densities, particularly
for  high-J  $  ^{12}$CO  transitions.   This is  why  some  of  their
brightness   temperature  line  ratios   are  unrealistic,   e.g.   CI
$(2-1)/(1-0)\sim  2.25$ (Table  1 in  Blain  et al.  2000), while  the
highest value  observed to date is  $\sim 1$, in M~82  (Stutzki et al.
1997) and the Orion KL region (Yamamoto~2001).

There  is  no reason  to  believe  that  the variety  of  environments
indicated  in Figure 5  will not  be also  present at  high redshifts.
Even if starburst environments  are more frequently encountered there,
the possibility of large  scale sub-thermal excitation as suggested by
recent measurements  in the extremely  red object (ERO)  and starburst
HR~10 at $\rm z\sim 1.45$  (Greve et al.  2003), cannot be discounted.
The  excitation turnover at  the frequencies  of the  CI J=1--0  and $
^{12}$CO  J=4--3 lines  implies that,  with  the exception  of the  CI
J=2--1 line, the gas mass tracing capability of higher frequency lines
($\nu >  492$ GHz) is suppressed except  in star-forming environments.
For essentially  the same reasons  molecular gas mass  estimates based
solely on  high-J $  ^{12}$CO lines observed  at high  redshifts carry
large  uncertainties  (factors   of  $\ga  $10).   Summarizing,  while
metal-rich gas  fueling star-formation is bound  to figure prominently
in high-J $  ^{12}$CO line emission at high redshifts,  it will not be
the  only gas  phase  present  and large  amounts  of colder,  diffuse
molecular gas  may be  lingering e.g.  ``in-between''  successive star
forming events beyond the current star-forming sites.

Here we  must mention that the  case for observing the  low $ ^{12}$CO
J=1--0, 2--1  lines remains  strong even if  the CI lines  can deliver
more flux  per H$_2$ mass for  e.g.  cool and  diffuse gas reservoirs.
This  is because  of the  current commissioning  of large  single dish
radio telescopes like  the Green Bank Telescope (GBT)  (tunnable up to
115  GHz),  and  future   instruments  as  the  expanded  VLA  (EVLA),
(continuously  tunnable below 45  GHz), and  a Square  Kilometer Array
(SKA)  equipped  with high  frequency  receivers  ($\sim 20-45$  GHz).
These will be able to conduct very sensitive observations of the low-J
$  ^{12}$CO transitions  at high  redshifts (Carilli  \&  Blain 2002),
bypass  the excitation  biases mentioned  previously,  and sensitively
trace molecular gas in the distant Universe.

\section{Deducing H$_2$ mass from CI line intensities: the uncertainties}

The  CI  lines  have  small  to moderate  optical  depths  ($\rm  \tau
_{CI}\sim 0.1-1$)  being, over  large scales, as  optically thin  as $
^{13}$CO (Ojha et el.  2001; Ikeda et al.  2002).  This property along
with their simpler 3-level  partition function and easy thermalization
makes  them  more straightforward  H$_2$  gas  mass  tracers than  the
usually optically thick $ ^{12}$CO lines.

For an optically thin line the emergent brightness  is 

\begin{equation}
\rm I_{\nu _{em}}=\frac{h\nu _{ul}}{4\pi} A_{ul} N_{u} \phi _{\nu_{em}}
\end{equation}

\noindent
where $\rm  N_u$ is the column  density of the upper  level, and $\phi
_{\nu _{em} }$ is the intrinsic line profile.  The received brightness
is  $\rm  I_{\nu }=(1+z)^{-3}  I_{\nu  _{em}}$  (see  e.g.  Ivison  et
al. 1996).  Thus  in terms of the usually  reported observable $\rm I=
\int  S_{\nu}  dV$ ($\rm  S_{\nu}=  \int  I_{\nu  } d\Omega$)  the  CI
beam-averaged column  density over the telescope beam  $\rm \Omega _b$
can be expressed as

\begin{equation}
\rm N_{CI}=\frac{4\pi (1+z)^3 I_{CI}}{h c A_{ul} Q_{ul} \Omega _b}.
\end{equation}

\noindent
where the $\rm Q_{ul}=N_u/N_{CI}<1$ excitation factor depends on ($\rm
 n$, $\rm T_{k}$),  and the radiation field (see  Appendix).  In terms
 of the gas mass contained  within $\rm \Omega _b$ the last expression
 becomes

\begin{equation}
\rm M_{A_b}=\frac{4\pi \mu  m_{H_2}}{hc A_{ul} X_{CI}} \left(\frac{D^2
_L}{1+z}\right) Q^{-1} _{ul} I_{CI},
\end{equation}

\noindent
where  we  used  $\rm  A_b  =(1+z)^{-4}  D^2  _L  \Omega  _b$  as  the
corresponding area  of the  beam at the  redshift of the  source, $\rm
X_{CI}=[C]/[H_2]$, and  $\mu =1.36$  corrects for the  mass in  He. In
astrophysical units,

\begin{eqnarray}
\rm \frac{M_{A_b}}{M_{\odot}}=4.92\times 10^{10} h ^{' -2}
\frac{(1+z-\sqrt{1+z})^2}{1+z}
\left[\frac{X_{CI}}{10^{-5}}\right]^{-1}
\rm \left[\frac{A_{ul}}{10^{-7} s^{-1}}\right]^{-1} Q_{ul} ^{-1}
\left[ \frac{I_{CI}}{Jy\ km\ s^{-1}}\right], 
\end{eqnarray}

\noindent
where we set $\rm H_{\circ}=100h^{'} \ km\ s^{-1} Mpc^{-1}$

For  optically thick  emission the  latter expression  yields  a lower
limit on the gas mass and remains valid for optical depths emerging in
radiatively decoupled ``cells'' much smaller than the emitting region,
but with $\rm  A_{ul}$ replaced by $\rm \beta  _{ul} A_{ul}$ ($\beta =
(1-e^{-\tau})/\tau $: photon escape probability).  In molecular clouds
where $\rm  \tau _{CI}\la 1$, an  unknown value of $\rm  \beta _{ul} $
introduces  an  uncertainty  of  $\la  60\%$.   A  greater  source  of
uncertainty  stems   from  the  assumed   excitation  conditions  that
determine $\rm Q_{ul}$.   For high redshift objects where  only one or
two lines are  usually detected there can be  no practical constraints
on the  gas excitation  and the minimal  assumption of LTE  is usually
made  in  order   to  deduce  $\rm  M(H_2)$.   For   typical  $\rm  n=
(100-5\times  10^4)\ cm^{-3}$  and  $\rm T_k=(20-60)\  K$  it is  $\rm
Q_{10}/Q^{(LTE)}   _{10}\sim  0.35-1   $  and   $\rm  Q_{21}/Q^{(LTE)}
_{21}\sim 0.15-1$.  These uncertainties can  be larger if there are no
available temperature  constraints from e.g. measurements  of the dust
continuum.  Predictably $\rm Q_{21}$ has the largest variation, making
the CI  J=2--1 transition the  more uncertain H$_2$ mass  estimator of
the  two CI  lines.   Nevertheless the  $  ^{12}$CO J=7--6  line at  a
similar  frequency  (806  GHz)  is  much  more  sensitive  to  ambient
conditions,  and  hence an  even  more  uncertain  H$_2$ mass  tracer.
Similarly  the CI  J=1--0 with  a  smaller $\rm  Q_{10}$ variation  is
tracing H$_2$ gas  more reliably than the $  ^{12}$CO J=4--3 (461 GHz)
line even though the latter can be significantly stronger.

In  deducing $\rm  M(H_2)$ from  CI  lines we  assumed CI  distributed
throughout  a   typical  molecular  cloud  but   also  no  significant
``contamination'' from any  other gas phase.  This is  indeed the case
since carbon in HI gas  is mostly ionized with e.g. $\rm [CI]/[CII]\la
3\times 10^{-3}$ in  the CNM phase (Wolfire et  al.  2003).  The small
CI line optical depths are an advantage when estimating H$_2$ mass but
they also  add $\rm X_{CI}$ to  the list of  uncertainties.  Given the
variety  of  processes  affecting  [CI]/[$  ^{12}$CO]  and  thus  $\rm
X_{CI}$, it seems  that the latter can vary  widely.  However {\it the
very same processes also  affect $\rm X_{CO}=[ ^{12}CO]/[H_2]$}, which
is nevertheless found to  be relatively constant ($\sim 10^{-4}$) over
large  scales for  clouds with  $\rm A_v\ga  2 $  (e.g.   van Dishoeck
1992).   This  is  in  accordance  with  the  constant  $\rm  N(CI)/N(
^{12}CO)\sim 0.1-0.2$ found by large  scale CI, $ ^{12}$CO, $ ^{13}$CO
J=1--0  surveys (e.g.   Ikeda  et  al 2002)  and  suggests a  dominant
physical process that along with turbulent diffusion sets a relatively
constant [C]/[$ ^{12}$CO] throughout a  GMC.  However, as was the case
for $ ^{12}$CO, establishing average $\rm X_{CI}$ values will probably
be  done observationally, with  large surveys  conducted over  a broad
range  of  conditions and  discerning  any  general  trends of  [C]/[$
^{12}$CO]  (and  thus  $\rm   X_{CI}$)  in  active  and/or  metal-poor
environments.

\section{Observational prospects at high redshifts}

The potential advantage  of the CI lines as H$_2$  mass tracers can be
fully realized only  at high redshifts.  For $\rm  z_{10}\ga 1$ the CI
J=1--0  line is redshifted  to $\la  250$ GHz  where a  combination of
existing large-aperture instruments  (single dish and interferometers)
and  a  much   improved  atmospheric  transmission  enables  sensitive
observations.  For  the J=2--1 transition the  aforementioned limit is
$\rm  z_{21}\ga 2.25$,  a typical  redshift of  the new  population of
sub-mm  bright,  starburst  galaxies  (Chapman  et  al.   2003).   The
commissioning  of interferometers  in excellent  sites, namely  SMA on
Mauna Kea, Hawaii  and ALMA in Llano de  Chajnantor, Chile, will allow
routine sensitive  observations at sub-mm wavelengths.   In such sites
the  atmospheric   transmission  even   at  $\sim  345$   GHz  remains
exceptional most of the time ($\sim 80\%$ at zenith, except around the
325  GHz atmospheric  absorption  feature), and  the  CI lines  become
accessible to sensitive measurements for $\rm z_{10} \ga 0.4$ and $\rm
z_{21}\ga 1.3$.  Particularly interesting  is the case when {\it both}
CI  lines become  accessible at  lower frequencies,  thus  allowing an
estimate of gas temperatures from their intensity ratio (assuming LTE)
independently  of  those  based   on  dust  continuum  or  $  ^{12}$CO
transitions.

Currently all  high redshift  objects detected in  $ ^{12}$CO  are the
sites  of  large  scale  starbursts   and  are  good  targets  for  CI
observations, which hold the  promise of uncovering H$_2$ gas untraced
by  the high-J  $ ^{12}$CO  transitions.  Recently  an  opportunity to
drastically increase  their rather small numbers has  emerged with the
successful  spectroscopic  redshift  determination of  several  sub-mm
bright  starbursts at  high redshifts  (Chapman et  al.   2003).  Such
objects  are metal-rich  and  are  found to  be  prominent $  ^{12}$CO
emitters (e.g.   Neri et al.  2003).   The recent detection  of weak $
^{12}$CO J=3--2  in a Ly-break galaxy  at z=2.7 (Baker  et al.  2003),
with  its CI  J=2--1 line  redshifted  to $\sim  218$~GHz, offers  the
possibility of testing the potential  of CI lines as H$_2$ gas tracers
in metal-poor and modestly UV-intense objects.

 The  recent detection  of  massive gaseous  disks  at high  redshifts
(Genzel et al.   2003) are currently a highly  contentious issue since
their presence at such early epochs is hard to reconcile with standard
hierarchical  models of  galaxy  evolution.  The  much larger  spatial
distribution of the CI line emission with respect to high-J $ ^{12}$CO
transitions  in typical  disks  like  the Milky  Way,  {\it holds  the
promise of yielding a much better picture of disk kinematics and total
dynamical mass}  (by encompassing its  flat rotation curve  part), and
thus is an excellent~alternative.

There are currently  only two reported detections of  CI lines at high
 redshifts, namely in the Cloverleaf quasar (z=2.56) (Barvainis et al.
 1997; Weiss  et al.  2003),  and the IRAS galaxy  F10214+472 (z=2.28)
 (Brown \& Vanden Bout 1992)\footnote{This detection was under dispute
 but  it has  been recently  confirmed (Papadopoulos  et  al.  2004)}.
 Both are strongly lensed  systems and particularities of differential
 lensing  may be responsible  for their  very different  CI/$ ^{12}$CO
 flux density  ratios. Clearly CI  observations of  non-lensed high
 redshift objects are needed.

Interestingly  in  the  case  of  the Cloverleaf,  and  based  on  the
 assumption of a concomitant CI  emission, Weiss et al.  find an H$_2$
 gas mass similar to that deduced from the well-sampled dust continuum
 or from the multi-transition $ ^{12}$CO observations, lending support
 to the main argument of this work.

Finally,  the markedly  more difficult  CI observations  in  the local
Universe  remain valuable  since large  scale  CI, $  ^{12}$CO, and  $
^{13}$CO surveys in a variety  of environments are needed to establish
the range of  $\rm X_{CI}$ and ``cross-calibrate'' the  two methods of
estimating H$_2$ gas mass.   It is particularly interesting to observe
intensely star-forming  and/or metal-poor  objects in order  to search
for  [C]/[$  ^{12}$CO]   enhancements  and  its  possible  metallicity
dependence.

 \section{Conclusions}

In  this work  we examined  the prospects  of using  the two  lines of
atomic carbon at 492~GHz  (J=1--0) and 809 GHz (J=2--1) as alternative
H$_2$ mass tracers in galaxies.   Our conclusions can be summarized as
follows:

1. Non-equilibrium  chemical states,  turbulent diffusive  mixing, and
   cosmic  rays can  easily make  CI ubiquitous  throughout  a typical
   Giant Molecular Cloud.  This is now firmly supported by large scale
   CI, $ ^{12}$CO, $ ^{13}$CO surveys of such clouds in the Milky Way,
   which also find a constant average [C]/[$ ^{12}$CO] abundance ratio
   over the bulk of their mass.

2. In star  forming and/or dynamically  fast-evolving environments the
   processes responsible for  the ubiquity of CI in  GMCs will further
   enhance its abundance.  In  UV-intense and metal-poor objects large
   scale CII production diminishes the H$_2$-tracing capability of CI,
   which  nevertheless remains better  than that  of the  now severely
   dissociated $ ^{12}$CO.

3. In  terms of  emerging flux  density  as a  function of  excitation
  conditions CI lines have a potentially strong advantage with respect
  to $ ^{12}$CO  lines in tracing H$_2$ gas  mass under diffuse and/or
  cooler conditions as well as  in the metal poor environments of mild
  starbursts.   However  the  large  atmospheric absorption  at  their
  rest-frame frequencies  allows such  an advantage to  materialize at
  high redshifts ($\rm z\ga 1$).

4. The simpler 3-level structure of  the CI lines, their small optical
    depths  and  modest excitation  requirements  with  respect to  $
    ^{12}$CO lines of similar frequency make the associated H$_2$ mass
    estimates  more  straightforward.   However  the  usually  unknown
    (particularly   for  high   redshift   sources)  mean   excitation
    conditions introduce large  uncertainties in such estimates, which
    are reduced considerably if an  estimate of the gas temperature is
    available.

5. Concentrated  observational effort  in the  local  Universe remains
   critical.  Large  scale CI, $  ^{12}$CO, and $ ^{13}$CO  surveys of
   GMCs  in a  variety of  environments  will help  address issues  of
   optical depths, $\rm [CI]/[H_2] $ abundances and its variations, and
   ``cross-calibrate'' the two H$_2$ gas mass tracing techniques.

\section*{Acknowledgments}
We thank A.   Weiss for a detailed reading  of the original manuscript
and critical comments  that have greatly improved it.  Comments by the
anonymous referee are greatfuly acknowledged. P.P.P.  thanks Arjen van
der Wel  for fruitful  discussions and acknowledges  the support  of a
Marie Curie Individual Fellowship HPMT-CT-2000-00875.

\newpage

\clearpage

\begin{table}
\caption{CO, CI line ratios, typical LVG solutions}
\label{tab1}
\begin{center}
\begin{tabular}{l c c c c c l l}
\hline\hline
Environment$ ^{\rm a}$ & $\rm R_{21}$$ ^{\rm b}$ & $\rm R_{32}$$ ^{\rm b}$ &
$\rm K_{10}$$ ^{\rm c}$ & $\rm K_{21}$$ ^{\rm c}$ &
 $\rm R_{ci}$ ($\rm r_{21}$)$ ^{\rm d}$ & LVG solutions$ ^{\rm e}$ & References\\
\hline\hline
Very quiescent & 0.4 & 0.15  & 9  & ....
& 0.10  (0.22)  &  10, $\rm 10^3$, $\rm 10^{-5}$ (40) & 1, 2, 3, 4 \\
          &      &      &    &    &                & 10, $300$, $10^{-4}$ (60) &  \\
Quiescent & 0.6 & 0.30 & 5  & .... 
 & 0.15 (0.30) & 20, 300, $10^{-4}$ (40) & 5, 6, 7, 8 \\
          &      &      &    &    &                &  10, $10^3$, $3\times 10^{-5}$ (60) & \\
Dynamically active & 0.95 & .... &  10 &  10 &
0.10 (0.22) & 60, $10^3$, $10^{-5}$ (25) & 9, 10, 11, 12\\
Starburst & 0.95 & 0.65 & 15 & 15
 & 0.15-0.30 (0.5) & 30, 300, $3\times 10^{-6}$ (40)$ ^{\rm f}$ & 13, 14, 15, 16, 17 \\
          &      &      &    &    &                 & 60, $10^3$, $10^{-5}$ (40) & \\
Extreme starburst$ ^{\rm g}$ & 0.95 & 0.65 & 25 & 25 & 0.15-0.30 (0.5) & 70, $10^3$, $10^{-5}$ (60) 
& 13, 14 \\
\hline
\end{tabular}
\end{center}

\noindent
$ ^{a}$ Very  quiescent: M31  cold clouds,  Quiescent: Milky
Way  disk, Dynamically active:  Galactic Center,  Starburst: various
line   surveys  in  several   objects,  Extreme   starbursts:  ULIRG nuclei\\
\noindent
$^ {\rm b}$ The   $  ^{12}$CO  (2--1)/(1--0),  (3--2)/(1--0) brightness temperature ratios\\
\noindent
$^ {\rm c}$ The $ ^{12}$CO/$ ^{13}$CO brightness temperature ratios for J=1--0, 2--1\\
\noindent
$^ {\rm d}$ The $ ^{12}$CO/CI J=1--0 and $\rm r_{21}$=CI(2--1)/(1--0) brightness temperature ratios\\
\noindent
$^ {\rm e}$ LVG solutions: $\rm T_{k}(K)$, $\rm n(H_2)(cm^{-3})$,
$\rm X_{CO}/(dV/dr) ([km\ s^{-1} pc^{-1}]^{-1})$, ($\rm [ ^{12}CO/ ^{13}CO]=40, 60$),
 $\rm X_{CO}=[ ^{12}CO/H_2]$. In case of serious ``degeneracies'' typical
solutions from both parts of the  parameter space are displayed (e.g. in starbursts).\\
\noindent
$ ^{f}$ Best solutions occur for $\rm [ ^{12}CO/ ^{13}CO]=40$\\
\noindent
$ ^{g}$ Indicative ratios only, large range of values observed 
 we adopt the $\rm R_{32}$ and CI values from starbursts.

\noindent
1: Israel,  Tilanus, \& Baas 1998, 2:  Allen et al.  3:  Fixsen et al.
1999 ({\it COBE}, outer Galaxy), 4: Wilson 1997 (M 33 clouds), 5: Disk
cloud surveys  by Chiar  et al.  1994;  Hasegawa 1997; Sakamoto  et al
1997;  Falgarone et  al.  1998;  Sanders et  al. 1993  (for  CO 3--2);
Digel, de  Geus, \& Thaddeus 1994,  6: Disk cloud  surveys by Solomon,
Scoville, \&  Sanders 1979; Rickard \&  Blitz 1985; Polk  et al. 1988;
Falgarone et  al. 1998;  Wilson et  al. 1999, 7:  Fixsen et  al.  1999
(inner Galaxy); 8:  Gerin \& Philips 2000; 9:  Langer \& Penzias 1993;
10: Sawada et al.  2001; 11: Ojha et al. 2001; 12: Fixsen et al.  1999
(Galactic  center); 13:  Braine \&  Combes 1992,  Aalto et  al.  1995,
Papadopoulos \& Seaquist 1998; 14:  Devereux et al.  1994; 15: Stutzki
et al.  1997; Yamamoto et al.   2001; 16: Israel, White, \& Baas 1995,
Petitpas \& Wilson  1998; Israel \& Baas 2002, 17:  Stark et al. 1997;
Bolatto et al. 2000a, 2000b

\end{table}

\clearpage

\appendix

\section{The 3-level system and the CI  lines}

The rate  equations for a 3-level  system in steady  state under the
 assumption of $\rm A_{20}/A_{21}\ll  1$, $\rm A_{20}/A_{10}\ll 1$, as
 is the case for the CI fine structure levels, are

\begin{eqnarray}
\rm \frac{dn_2}{dt} = 
  B_{12} J_{\nu _{21}} n_1-\left(B_{21} J_{\nu _{21}} + A_{21}\right)n_2 +
  \left[C_{12} n_1 +  C_{02} n_0 -  C_{21} n_2 -  C_{20} n_2\right]=0, 
\end{eqnarray}

\begin{eqnarray}
\rm \frac{dn_1}{dt} =
\left(B_{21} J_{\nu _{21}} +A_{21}\right) n_2 + B_{01} J_{\nu _{10}} n_0 - 
\left(A_{10}+B_{10} J_{\nu _{10}} + B_{12} J_{21} \right) n_1 + \nonumber \\
\rm \left[C_{21}n_2+C_{01}n_0-C_{12}n_1-C_{10}n_1\right]=0,
\end{eqnarray}

\noindent
where $\rm J_{\nu}=2h\nu ^3/c^2 \left[exp(h\nu/kT_b)-1\right]^{-1}$ is
the  background radiation field  (assumed isotropic  and with  no line
contribution),  $\rm   C_{ul}=n  \gamma   _{ul}$  are  the   rates  of
collisional de-excitation  ($\rm u\rightarrow l$) related  to those of
the reverse  process through the  principle of detailed  balanced $\rm
C_{lu}/C_{ul} =  (g_{u}/g_{l}) exp(-h\nu _{ul}/kT_{k})$.   The term in
the square brackets is the net collisional population/de-population of
the levels, and the rest are the stimulated and spontaneous processes.

 The stimulated excitation  and de-excitation coefficients are related
as  $\rm g_u  B_{ul}=  g_l  B_{lu}$ and  to  the spontaneous  emission
coefficient $\rm A_{ul}=(2h \nu ^3 _{ul}/c^2) B_{ul}$.  The latter can
be  used  to  demonstrate  that  $\rm  B_{20}/B_{21}\ll  1$  and  $\rm
B_{20}/B_{10}\ll  1$, also  assumed in  the previous  equations. After
some reordering  and replacing all  the $\rm B_{ik}$ factors  we get

\begin{eqnarray}
\rm \left(C_{21}+C_{20}+f_{21}A_{21}\right)n_2-\left[C_{12}+g_2/g_1\left(f_{21}-1\right)A_{21}\right]n_1-
C_{02}n_0=0,
\end{eqnarray}

\begin{eqnarray}
\rm \left(C_{21}+f_{21} A_{21}\right)n_2 -\left[C_{10}+C_{12}+f_{10}A_{10}
+g_2/g_1\left(f_{21}-1\right)A_{21}\right]n_1+\nonumber \\
\rm \left[C_{01}+g_1/g_0\left(f_{10}-1\right)A_{10}\right] n_0=0,
\end{eqnarray}

\noindent
with $\rm  f=exp(h\nu/kT_b)/[exp(h\nu/kT_b)-1]$.   Equations A3,  A4
along with the ``closure'' relation

\begin{equation}
\rm n_2+n_1+n_0= n_{CI},
\end{equation}

\noindent
are used  to solve  for the level  populations. We find  the following
expressions,

\begin{eqnarray}
\rm \frac{n_1/n_{CI}}{K^{-1}}=1+f_{21}\left(1+\frac{C_{20}}{C_{21}}\right)\frac{n_{21}}{n}
+\frac{C_{01}}{C_{02}}\left(1+\frac{C_{20}}{C_{21}}\right)\left(1+f_{21}\frac{n_{21}}{n}
\right)\times \nonumber \\
\rm \left[1+\frac{g_1}{g_0}\left(f_{10}-1\right)\left(\frac{C_{12}+C_{10}}{C_{01}}\right)
\frac{n_{10}}{n} \right],
\end{eqnarray}

\begin{eqnarray}
\rm \frac{n_2/n_{CI}}{K^{-1}}=\left(1+f_{10}\frac{n_{10}}{n}\right)\frac{C_{10}+C_{12}}{C_{21}}+
\frac{g_2}{g_1}\left(f_{21}-1\right)\left(1+\frac{C_{20}}{C_{21}}\right)\frac{n_{21}}{n}+
\frac{C_{10}}{C_{20}}\times \nonumber \\
\rm \left[1+\frac{g_2}{g_1}\left(f_{21}-1\right)\left(\frac{C_{21}+C_{20}}{C_{12}}\right) 
\frac{n_{21}}{n}\right]\times 
\left[1+\frac{g_1}{g_0}\left(f_{10}-1\right)\left(\frac{C_{10}+C_{12}}{C_{01}}\right)
\frac{n_{10}}{n}\right], 
\end{eqnarray}

\noindent
and

\begin{eqnarray}
\rm \frac{n_0/n_{CI}}{K^{-1}} = \frac{C_{10}}{C_{02}} \left(1+f_{21}\frac{n_{21}}{n}\right)
\left(1+\frac{C_{20}}{C_{21}}\right)\left[1+f_{10}\left(1+\frac{C_{12}}{C_{10}}\right)
\frac{n_{10}}{n}\right]+\frac{C_{10}}{C_{01}}\times \nonumber \\
\rm \left[1+\frac{g_2}{g_1}\left(f_{21}-1\right)\left(\frac{C_{20}+C_{21}}{C_{12}}\right)
\frac{n_{21}}{n}\right]
\end{eqnarray}

\noindent
It is  $\rm Q_{10}=n_1/n_{CI}$  and $\rm Q_{21}=n_2/n_{CI}$  (see main
text) while the critical density  of a transition $\rm u\rightarrow l$
is  $\rm  n_{ul}=A_{ul}/\left(\sum _{k}  \gamma  _{uk}\right) $  (e.g.
Jansen 1995), where  the sum is over all  collisional processes out of
the level  (u).  In  the literature often  only the  $\rm \gamma_{ul}$
coefficient or the  sum of just the downwards  rates is used resulting
to an  often serious  overestimate of $\rm  n_{ul}$.  For  the 3-level
atom it  is $\rm n_{21}= A_{21}/(\gamma _{20}+\gamma  _{21})$ and $\rm
n_{10}=A_{10}/(\gamma _{12}+\gamma _{10})$ ($\rm n_{10}>n_{21}$).  The
expression for K is given by

\begin{eqnarray}
\rm K=1+\left(1+\frac{C_{20}}{C_{21}}\right) \Bigg\{\left(f_{21}+\frac{g_2}{g_1}\left(f_{21}-1\right)\right)
\frac{n_{21}}{n}+\frac{C_{10}}{C_{02}}\left(1+f_{21}\frac{n_{21}}{n}\right)
\left(1+\frac{C_{01}}{C_{10}}\right)\times \nonumber \ \ \ \ \ \ \ \ \ \ \ \ \ \ \ \ \  \\ 
\rm \left[1+\left(f_{10}+\frac{g_1}{g_0}\left(f_{10}-1\right)\right)
\left(\frac{C_{10}+C_{12}}{C_{01}+C_{10}}\right)
\frac{n_{10}}{n}\right]\Bigg\}+\left(1+f_{10}\frac{n_{10}}{n}\right)
\frac{C_{12}+C_{10}}{C_{21}}+\frac{C_{10}}{C_{01}} \left(1+\frac{C_{01}}{C_{20}}\right)\times\nonumber \\
\rm \left[1+\frac{g_2}{g_1}\left(f_{21}-1\right)\left(\frac{C_{20}+C_{21}}{C_{12}}\right)
\frac{n_{21}}{n}\right]\times \left[1+\frac{g_1}{g_0}\left(f_{10}-1\right)
\left(\frac{C_{10}+C_{12}}{C_{20}+C_{01}}\right)
\frac{n_{10}}{n}\right].\ \ \ \ 
\end{eqnarray}

\noindent
 In the case $\rm n_{21}/n\rightarrow 0 $ (and thus also $\rm n_{10}/n\rightarrow 0$), the
last expressions yield

\begin{equation}
\rm \frac{n_1/n_{CI}}{K^{-1}}\rightarrow 1+\left(1+\frac{C_{20}}{C_{21}}\right)\frac{C_{01}}{C_{02}}
=1+\frac{C_{01}}{C_{02}}+\frac{C_{10}}{C_{12}}, 
\end{equation} 

\begin{equation}
\rm \frac{n_2/n_{CI}}{K^{-1}}\rightarrow \frac{C_{10}+C_{12}}{C_{21}} + \frac{C_{10}}{C_{20}}=
\frac{C_{12}}{C_{21}}\left(1+\frac{C_{01}}{C_{02}}+\frac{C_{10}}{C_{12}}\right),
\end{equation}

\noindent
and

\begin{equation}
\rm \frac{n_0/n_{CI}}{K^{-1}}\rightarrow \left(1+\frac{C_{20}}{C_{21}}\right)\frac{C_{10}}{C_{02}}+
\frac{C_{10}}{C_{01}}=\frac{C_{10}}{C_{01}}\left(1+\frac{C_{01}}{C_{02}}+\frac{C_{10}}{C_{12}}\right).
\end{equation}

\noindent
The last relations yield

\begin{equation}
\rm \frac{n_2}{n_1}=\frac{C_{12}}{C_{21}}=\frac{g_2}{g_1}e^{-E_{21}/kT_k}, \ \ and \ \ 
\frac{n_1}{n_0}=\frac{C_{01}}{C_{10}}=\frac{g_1}{g_0}e^{-E_{10}/kT_k},
\end{equation}

\noindent
which are the LTE values while the expression for K becomes

\begin{eqnarray}
\rm K\rightarrow 1+\left(1+\frac{C_{20}}{C_{21}}\right)\left(1+\frac{C_{01}}{C_{10}}\right)
\frac{C_{10}}{C_{02}}+\frac{C_{12}+C_{10}}{C_{21}}+\frac{C_{10}}{C_{01}}
\left(1+\frac{C_{01}}{C_{20}}\right)= \nonumber \\
\rm \left(1+\frac{C_{01}}{C_{02}}+\frac{C_{10}}{C_{12}}\right)
\left(1+\frac{C_{12}}{C_{21}}+\frac{C_{10}}{C_{01}}\right).
\end{eqnarray}

\noindent
Thus from Equations A10 and A14 it is

\begin{equation}
\rm \frac{n_1}{n_{CI}}=\frac{1}{1+\frac{C_{12}}{C_{21}}+\frac{C_{10}}{C_{01}}}=
\frac{g_1 e^{-E_1/kT_k}}{g_0+g_1 e^{-E_1/kT_k}+g_2 e^{-E_2/kT_k}},
\end{equation}

\noindent
which its  expected LTE value. In  the last derivations as  well as in
all previous computations we made use of the permutation relation $\rm
C_{02}  C_{21}  C_{10}=C_{20} C_{01}  C_{12}$  that  results from  the
detailed balance equations.

  In  the radiation-dominated  limit $\rm  n_{10}/n\rightarrow \infty$
(and thus also $\rm n_{21}/n\rightarrow \infty$) we eventually deduce

\begin{eqnarray}
\rm \frac{n_1}{n_{CI}}\rightarrow \frac{g_1/g_0 f_{21}\left(f_{10}-1\right)}{f_{21}\left[f_{10}+g_1/g_0
\left(f_{10}-1\right)\right]+g_2/g_0\left(f_{10}-1\right)\left(f_{21}-1\right)}= \nonumber \\
\rm \frac{g_1 e^{-E_1/kT_b}}{g_0+g_1 e^{-E_1/kT_b}+g_2 e^{-E_2/kT_b}},
\end{eqnarray}

\noindent
and

\begin{equation}
\rm \frac{n_2}{n_1}\rightarrow \frac{g_2}{g_1}\left(\frac{f_{21}-1}{f_{21}}\right)=
\frac{g_2}{g_1} e^{-E_{21}/kT_b}, \ \ \ \ 
\rm \frac{n_1}{n_0}\rightarrow \frac{g_1}{g_0}\left(\frac{f_{10}-1}{f_{10}}\right)=
 \frac{g_1}{g_0} e^{-E_{10}/kT_b},
\end{equation}

\noindent
which are the values expected for levels fully regulated by the background
radiation field.

The case where $\rm n_{21}\gg n_{10}$ allows partial thermalization of
the  3-level  system  since  there  is a  density  domain  where  $\rm
n_{10}/n\rightarrow 0$  while $\rm n_{21}/n$ remains  finite and could
be even $\la 1$. We then find the limit

\begin{equation}
\rm \frac{n_1}{n_0}\rightarrow \frac{g_1}{g_0} e^{-E_{10}/kT_k} \Bigg\{
\frac{1+\left(1+\frac{C_{20}}{C_{21}}\right)\frac{C_{01}}{C_{02}}
\left[1+f_{21}\left(1+\frac{C_{02}}{C_{01}}\right)\frac{n_{21}}{n}\right]}
{1+\left(1+\frac{C_{20}}{C_{21}}\right)\frac{C_{01}}{C_{02}}
\left[1+f_{21}\left(1+g_2/g_1\left(\frac{f_{21}-1}{f_{21}}\right)\frac{C_{20}}{C_{10}}\frac{n_{21}}{n}
\right)\right]}\Bigg\}.
\end{equation}

\noindent
Interestingly this still  deviates from its LTE value  unless two more
conditions  are   met,  namely  a)  $\rm   f_{21}\sim  1$  (negligible
background at the $\nu = \nu _{21}$ frequency), which yields

\begin{equation}
\rm \frac{n_1}{n_0}\rightarrow \frac{g_1}{g_0} e^{-E_{10}/kT_k} \Bigg\{
\frac{1+\left(1+\frac{C_{20}}{C_{21}}\right)\frac{C_{01}}{C_{02}}
\left[1+\left(1+\frac{C_{02}}{C_{01}}\right)\frac{n_{21}}{n}\right]}
{1+\left(1+\frac{C_{20}}{C_{21}}\right)\frac{C_{01}}{C_{02}}\left(1+\frac{n_{21}}{n}\right)}\Bigg\},
\end{equation}

\noindent
and b) $\rm  C_{02}/C_{01}\ll 1$ which then reduces  the factor in the
curly   brackets  to   unity.   These   conditions  ensure   that  the
non-thermalized J=2  level has  negligible interaction with  the lower
two  either radiatively  or  collisionally. Condition  (a) is  usually
satisfied for optically thin mm/sub-mm lines but condition (b) imposes
a temperature constraint, namely

\begin{equation}
\rm T_{k}\ll \left(E_{21}/k\right)
\left[ln\left(\frac{g_2 C_{20}}{g_1 C_{10}}\right)\right]^{-1}.
\end{equation}

In the  case of  the optically  thin CI lines  and since  usually $\rm
T_{k}\gg  T_{b}\sim  2.7$ K  we  use  the  negligible radiation  field
approximation  ($\rm f_{21}\sim  f_{10}\sim 1$)  to estimate  the $\rm
Q_{10}$ and $\rm Q_{21}$ factors, which are then are given by

\begin{equation} 
\rm Q_{10}=K^{-1}\Bigg\{1+\left(1+\frac{C_{20}}{C_{21}}\right)\frac{C_{01}}{C_{02}}
\left[1+\left(1+\frac{C_{02}}{C_{01}}\right)\frac{n_{21}}{n}\right]\Bigg\},
\end{equation}

\begin{equation}
\rm Q_{21}=K^{-1}\left[\frac{C_{10}}{C_{20}}+\frac{C_{10}+C_{12}}{C_{21}}\left(
1+\frac{n_{10}}{n}\right)\right].
\end{equation}

\noindent
Using the detailed balance equations these become

\begin{equation}
\rm Q_{10}=K^{-1}\Bigg\{1+\frac{g_1}{g_2}\left(1+\frac{C_{20}}{C_{21}}\right)
\frac{C_{10}}{C_{20}} e^{E_{21}/kT_k}\left[1+\left(1+\frac{g_2}{g_1}
\frac{C_{20}}{C_{10}}e^{-E_{21}/kT_k}\right)\frac{n_{21}}{n}\right]\Bigg\}
\end{equation}

\begin{equation}
\rm Q_{21}=K^{-1}\left[\frac{C_{10}}{C_{20}}+\left(\frac{C_{10}}{C_{21}}+
\frac{g_2}{g_1}e^{-E_{21}/kT_k}\right)\left(1+\frac{n_{10}}{n}\right)\right].
\end{equation}

\noindent
In the weak radiation field limit the expression for K becomes

\begin{eqnarray}
\rm K=1+\left(1+\frac{C_{20}}{C_{21}}\right)\Bigg\{\frac{n_{21}}{n}+\frac{C_{10}}{C_{02}}
\left(1+\frac{C_{01}}{C_{10}}\right) \left[1+\frac{C_{10}}{C_{01}}\left(\frac{1+C_{12}/C_{10}}{
1+C_{10}/C_{01}}\right)\frac{n_{10}}{n}\right]\times \nonumber \\
\rm \left(1+\frac{n_{21}}{n}\right)\Bigg\}+\frac{C_{12}}{C_{21}}
\left(1+\frac{C_{10}}{C_{12}}\right)\left(1+\frac{n_{10}}{n}\right)+\frac{C_{10}}{C_{01}}\left(1+
\frac{C_{01}}{C_{20}}\right),\ \ \ \ \ \
\end{eqnarray}

\noindent
or after some rearrangement of the terms and use of the detailed balance relations

\begin{eqnarray}
\rm K=1+\left(1+\frac{C_{20}}{C_{21}}\right)\Bigg\{\frac{n_{21}}{n}+\frac{g_0}{g_2}e^{E_2/kT_k}
\left(1+\frac{g_1}{g_0}e^{-E_1/kT_k}\right)\frac{C_{10}}{C_{20}}
G(n,T_k)\left(1+\frac{n_{21}}{n}\right)\Bigg\}+ \nonumber \\
\rm \frac{g_2}{g_1}e^{-E_{21}/kT_k}\left(1+\frac{g_1}{g_2}\frac{C_{10}}{C_{21}}e^{E_{21}/kT_k}\right)
\left(1+\frac{n_{10}}{n}\right)+\frac{g_0}{g_1}e^{E_1/kT_k}
\left(1+\frac{g_1}{g_0}\frac{C_{10}}{C_{20}}e^{-E_1/kT_k}\right), \ \ \ \ \ \ \
\end{eqnarray}

\noindent
where we have set

\begin{equation}
\rm G(n, T_k)=1+\frac{g_0}{g_1} e^{E_1/kT_k} \left(\frac{1+\frac{g_2}{g_1}
\frac{C_{21}}{C_{10}}e^{-E_{21}/kT_k}}{1+\frac{g_0}{g_1}e^{E_{1}/kT_k}}\right)\frac{n_{10}}{n}
\end{equation}

The CI J=1--0, J=2--1 hyperfine lines have degeneracy factors given by
$\rm  g_{J}=2J+1$,   and  energy  levels  $\rm   E_1/k=23.6$  K,  $\rm
E_2/k=62.4$ K.  Their Einstein coefficients are $\rm A_{10}=7.93\times
10^{-8}   s^{-1}$,  $\rm   A_{21}=2.68\times  10^{-7}   s^{-1}$,  $\rm
A_{20}=2\times  10^{-14}  s^{-1}$, and  their  collisional rates  $\rm
\gamma   _{10}=1.3\times  10^{-10}  \   cm^3\  s^{-1}$,   $\rm  \gamma
_{21}=7.8\times 10^{-11}\  cm^3\ s^{-1}$, $\rm  \gamma _{20}=2.0\times
10^{-10}  \ cm^3  \ s^{-1}$  with a  very weak  temperature dependance
(Zmuidzinas et  al.  1988 and  references therein). Thus  the critical
densities as defined previously are $\rm n_{21}=964\ cm^{-3}$ and $\rm
n_{10}=A_{10}/(\gamma  _{10}+\gamma _{12})=n_{10}(0) \left[1+(g_2/g_1)
e^{-E_{21}/kT_k}  \gamma _{21}/\gamma _{10}\right]^{-1}$.   The latter
varies from $\rm \sim 600\ cm ^{-3}$ ($\rm T_k=0$ K) to $\rm \sim 300\
cm ^{-3}$ (high temperature limit).

\end{document}